\documentclass[preprint,10pt,3p]{elsarticle}
\usepackage{graphicx,latexsym,amssymb, epsfig}
\usepackage{multirow,amsmath,array,booktabs}
\usepackage{subfigure}
\usepackage{tabularx}
\usepackage{color}
\usepackage{bm}
\usepackage[figuresright]{rotating}
\usepackage[section]{placeins}
\usepackage{slashed}
\usepackage{graphicx}
\usepackage{tikz}

\newcolumntype{R}{>{\raggedleft\arraybackslash}X}

\newcommand{\nc}{\newcommand}       
\nc{\vc}[1] {\mbox{\boldmath $#1$}} 
\nc{\del}       {\partial}              
\nc{\bra}       {\langle}               
\nc{\ket}       {\rangle}               
\nc{\bras}[1]   {\langle #1|}           
\nc{\kets}[1]   {|#1\rangle}            
\nc{\mapleft}[1]{           
 \smash{\mathop{\,          %
  \hbox to 1.5cm{\rightarrowfill}\, }\limits_{#1}}}
\nc{\beq}     {\begin{eqnarray}} \nc{\eeq}    {\end{eqnarray}}
\nc{\nn}      {\\\nonumber} \nc{\vs}      {\vspace{-0.275cm}}
\nc{\fra}    {\frac{1}{2}}
\nc{\mb}        {\mathbf}

\begin{document}

\title{Nuclear matter properties with nucleon-nucleon forces up to fifth order \\ in the chiral expansion}

\author [NKU]{Jinniu Hu\fnref{info1}}
\author [TJU]{Ying Zhang}
\author [RBU]{Evgeny Epelbaum}
\author [JUL,BNU]{Ulf-G. Mei\ss ner}
\author [PKU,BUU,SU]{Jie Meng}
\address[NKU]{School of Physics, Nankai University, Tianjin 300071, China}
\address[TJU]{Department of Physics, School of Science, Tianjin University, Tianjin 300072, China}
\address[RBU]{Institut f\"{u}r Theoretische Physik II, Ruhr-Universit\"{a}t Bochum, D-44780 Bochum, Germany}
\address[JUL]{Institut f\"{u}r Kernphysik, Institute for Advanced Simulation and J\"{u}lich Center
for Hadron Physics, Forschungszentrum J\"{u}lich, D-52425 J\"{u}lich, Germany}
\address[BNU]{Helmholtz-Institut f\"{u}r Strahlen- und Kernphysik and Bethe Center for Theoretical Physics, Universit\"{a}t
	Bonn, D-53115 Bonn, Germany}
\address[PKU]{School of Physics and State Key Laboratory of Nuclear Physics and Technology, Peking University, Beijing 100871, China}
\address[BUU]{School of Physics and Nuclear Energy Engineering, Beihang University, Beijing 100191, China}
\address[SU]{Department of Physics, University of Stellenbosch, Stellenbosch 7602, South Africa}

\fntext[info1]{hujinniu@nankai.edu.cn}


\date{\today}
\begin{abstract}
The properties of nuclear matter are studied using state-of-the-art nucleon-nucleon forces up to fifth order in chiral effective field theory. The
equations of state of symmetric nuclear matter and pure neutron matter are calculated in the framework of the  Brueckner-Hartree-Fock theory. 
We discuss in detail the convergence pattern of the chiral expansion and the regulator dependence of the calculated equations of 
state and provide an estimation of the truncation uncertainty. For all employed values of the regulator, the fifth-order chiral two-nucleon potential 
is found to generate nuclear saturation properties similar to the available phenomenological high precision potentials.  
We also extract the symmetry energy of nuclear matter, which is shown to be quite robust with respect to the chiral order and the value of the regulator. 
\end{abstract}



\maketitle

\section{Introduction}
\label{sec:intro}

The nuclear force,  a residual strong force between colorless nucleons, lies at the very heart of nuclear physics. Enormous progress has been made towards its quantitative understanding since the seminal work by Yukawa on the one-pion-exchange mechanism, which has been published more than eight decades ago~\cite{yukawa35}. Already in the fifties of the last century, Taketani {\it{et al.}} have pointed out that the range of nucleon-nucleon ($NN$) potential can be divided into three distinct regions~\cite{taketani51}. While the long-distance interaction is dominated by one-pion exchange, the two-pion exchange mechanism plays an important role in the intermediate region of $r \sim 1 \ldots 2~$fm. Multi-pion exchange interactions are most essential in the core region. After the discovery of heavy mesons, the $NN$ potential was successfully modeled using the one-boson-exchange (OBE) picture~\cite{signell69,erkelenz74} with multi-pion exchange potentials being effectively parametrized by single exchanges of heavy mesons like $\sigma$-, $\omega$- and, $\rho$-mesons. With a fairly modest number of adjustable parameters, the OBE potential models such as the Bonn~\cite{machleidt87,machleidt89} and Nijmegen 93~\cite{stoks94} models were able to achieve a semi-quantitative description of $NN$ scattering data. Furthermore, based on the general operator structure of the two-nucleon interaction in coordinate space, a phenomenological $NN$ potential model was also developed by the Argonne group~\cite{wiringa84}.  In the 1990s, high-precision charge-dependent $NN$ potential models such as 
e.g.~the Reid93 and Nijmegen I, II~\cite{stoks94}, AV18~\cite{wiringa95} and the CD Bonn~\cite{machleidt01} potentials have been developed, which describe the available proton-proton and neutron-proton elastic scattering data with $\chi^2/$datum$\sim1$. 

While phenomenologically successful, the above mentioned
high-precision $NN$ potentials have no clear relation 
to quantum chromodynamics (QCD), the underlying theory of the strong interactions. Further, they do not provide a straightforward way to 
generate consistent and systematically improvable many-body forces and exchange currents and do not allow to estimate the 
theoretical uncertainty. In this sense, a more promising and systematic approach to nuclear forces and current operators 
has been proposed by Weinberg in the framework of chiral effective field theory (EFT) based on the most general 
effective chiral Lagrangian constructed in harmony with the symmetries of QCD \cite{weinberg90,weinberg91,weinberg92}. 
The first quantitative studies of $NN$ scattering up to
next-to-next-to-leading order (N$^2$LO) in the chiral expansion 
have been carried out by Ord\'o\~nez {\it{et al.}} \cite{ordonez94,ordonez96} using  time-ordered 
perturbation theory, see also \cite{epelbaum98,epelbaum00} where the calculations were done using the method of unitary transformations. 
In the early 2000s, the  $NN$ potential has been worked out to fourth order in the chiral 
expansion (N$^3$LO) by Epelbaum, Gl\"ockle and Mei{\ss}ner \cite{epelbaum05} and 
by Entem and Machleidt \cite{entem03} 
based on the expressions for the pion exchange contributions derived by Kaiser \cite{Kaiser:1999jg,Kaiser:1999ff,Kaiser:2001pc}. 
The corresponding three- and four-nucleon forces have also been worked
out to N$^3$LO \cite{Ishikawa:2007zz,Bernard:2007sp,Bernard:2011zr,Epelbaum:2006eu,Epelbaum:2007us}, see \cite{Epelbaum:2008ga,machleidt11}
for review articles and \cite{Krebs:2012yv,Krebs:2013kha,Girlanda:2011fh} for calculations beyond N$^3$LO. 
Recently, fifth- (N$^4$LO) and even some of the sixth-order contributions to the two-nucleon force 
have been worked out in \cite{entem15,Entem:2015xwa}, and a new
generation of chiral $NN$ potentials up to N$^4$LO utilizing a local coordinate-space 
regulator for the long-range terms has been introduced in \cite{epelbaum15a,epelbaum15b}. In parallel, a novel simple approach 
for estimating the theoretical uncertainty from the truncation of the chiral expansion has been proposed in \cite{epelbaum15a} and
successfully validated for two-nucleon observables \cite{epelbaum15a,epelbaum15b}. The algorithm makes use of the 
explicit knowledge of the contributions to an observable of interest at various orders in the chiral expansion without relying 
on cutoff variation. The new state-of-the-art $NN$ potentials confirm a good convergence of the chiral expansion for 
nuclear forces and lead to accurate description of Nijmegen phase shifts \cite{stoks93}. For related recent developments 
see Refs.~\cite{Gezerlis:2013ipa,Piarulli:2014bda}. 

Currently, work is in progress by the recently established 
Low Energy Nuclear Physics International Collaboration (LENPIC)~\cite{lenpic} 
towards including the consistently regularized 
three-nucleon force (3NF) at N$^3$LO in ab initio calculations of light- and medium-mass nuclei. In parallel, the novel chiral 
$NN$ potentials have been tested in nucleon-deuteron elastic scattering and properties of 
$^3$H, $^4$He, and $^6$Li \cite{binder16} and selected electroweak processes \cite{Skibinski:2016dve}, 
where special focus has been put on estimating the theoretical uncertainty at each order of the expansion.
These studies have revealed the important role of the 3NF, whose expected contributions to various bound and scattering 
state observables appear to be in good agreement with the expectation
based on the power counting. 

Light- and medium-mass nuclei can nowadays be studied using various ab initio methods such as the Green's function Monte Carlo method~\cite{carlson15}, 
the self-consistent Green's function method~\cite{dickhoff04}, the coupled-cluster approach~\cite{hagen14a}, 
nuclear lattice simulations~\cite{lee09,epelbaum14,Meissner:2015una} or the no-core-shell model~\cite{barrett13}, see also Ref.~\cite{shen16} 
for a first application of the relativistic Brueckner-Hartree-Fock theory to finite nuclei. 
Infinite nuclear matter has also been widely studied based on various versions of the chiral potentials 
using  e.g. the quantum Monte Carlo approach \cite{Gezerlis:2013ipa}, 
self-consistent Green's function method~\cite{carbone13,carbone14}, the coupled-cluster method~\cite{hagen14}, many-body perturbation theory~\cite{drischler14}, functional renormalization group (FRG) method \cite{drews15,drews16} and the Brueckner-Hartree-Fock (BHF) theory~\cite{li06,machleidt10}.
Recently, Sammarruca {\it{et al.}} have discussed the convergence of chiral EFT in infinite nuclear matter 
using the nonlocal $NN$ potentials up to N$^3$LO \cite{entem03} and
including the 3NF at the N$^2$LO (i.e.~$Q^3$) level 
\cite{sammarruca15}. Fairly large deviations between the results at
different chiral orders as compared with the spread in predictions
due to the employed cutoff variation have been reported in that
paper. This suggests that cutoff variation does not represent a
reliable approach to uncertainty quantification, which is fully in
line with the conclusions of \cite{epelbaum15a}.  
Regulator artifacts in uniform matter have also been addressed in Ref.~\cite{Dyhdalo:2016ygz}.

In this letter we calculate, for the first time, the properties of symmetric nuclear matter (SNM) and pure neutron matter (PNM) 
based  the latest generation of chiral $NN$ potentials 
up to N$^4$LO of Refs.~\cite{epelbaum15a,epelbaum15b} using the BHF
theory. The purpose of our study is twofold. First, we explore the
suitability of the most recent generation of the chiral forces for
microscopic description of the equation of state (EOS) of SNM and
PNM. Second, by performing an error analysis
along the lines of Refs.~\cite{epelbaum15a,epelbaum15b,binder16}
without relying on cutoff variation, we estimate the theoretical
accuracy in the description of the nuclear EOS achievable at various
orders of the chiral expansion. 
Our paper is organized as follows. 
In section \ref{sec:BHF} we briefly outline our calculation approach based on the BHF theory. The results of our calculations 
are presented in section \ref{sec:Results} for all available cutoff values, while the theoretical uncertainty from the truncation of the 
chiral expansion is quantified in  section \ref{sec:Errors}. Finally,  the main conclusions of our paper 
are summarized in section \ref{sec:Summary}.

\section{Brueckner-Hartree-Fock theory}
\label{sec:BHF}

In the BHF theory of nuclear matter, the underlying $NN$ potential, determined by the $NN$ scattering data, 
is replaced by an effective $NN$ interaction, i.e. the $G$-matrix, which can be calculated by solving the  
Bethe-Goldstone equation~\cite{li06,baldo07},
\beq
G[\omega,\rho]=V+\sum_{k_a,k_b>k_F}V\frac{|k_ak_b\ket\bra k_ak_b|}{\omega-e(k_a)-e(k_b)+i\epsilon}G[\omega,\rho],
\eeq
where $V$ is the underlying $NN$ potential provided by chiral EFT,  $\rho$ is the nucleon number density, and $\omega$ the starting energy. The single-particle energy is
\beq
e(k)=e(k;\rho)=\frac{k^2}{2m}+U(k,\rho).
\eeq

The continuous choice for the single-particle potential $U(k,\rho)$ used in the present BHF theory~\cite{baldo07} has the form 
\beq
U(k;\rho)={\rm Re} \sum_{k'<k_F}\bra kk'|G[e(k)+e(k');\rho]|kk'\ket_a,
\eeq
where the subscript $a$ indicates antisymmetrization of the matrix elements. These coupled equations are 
solved in a self-consistent way. Finally, in the BHF theory, we obtain the energy per nucleon as
\beq\label{et}
\frac{E}{A}=\frac{3}{5}\frac{k^2_F}{2m}+\frac{1}{2\rho}{\rm Re} \sum_{k,k'<k_F}\bra kk'|G[e(k)+e(k');\rho]|kk'\ket_a.
\eeq

\section{Results}
\label{sec:Results}

\begin{figure*}[t]
	\includegraphics[width=1.0\textwidth]{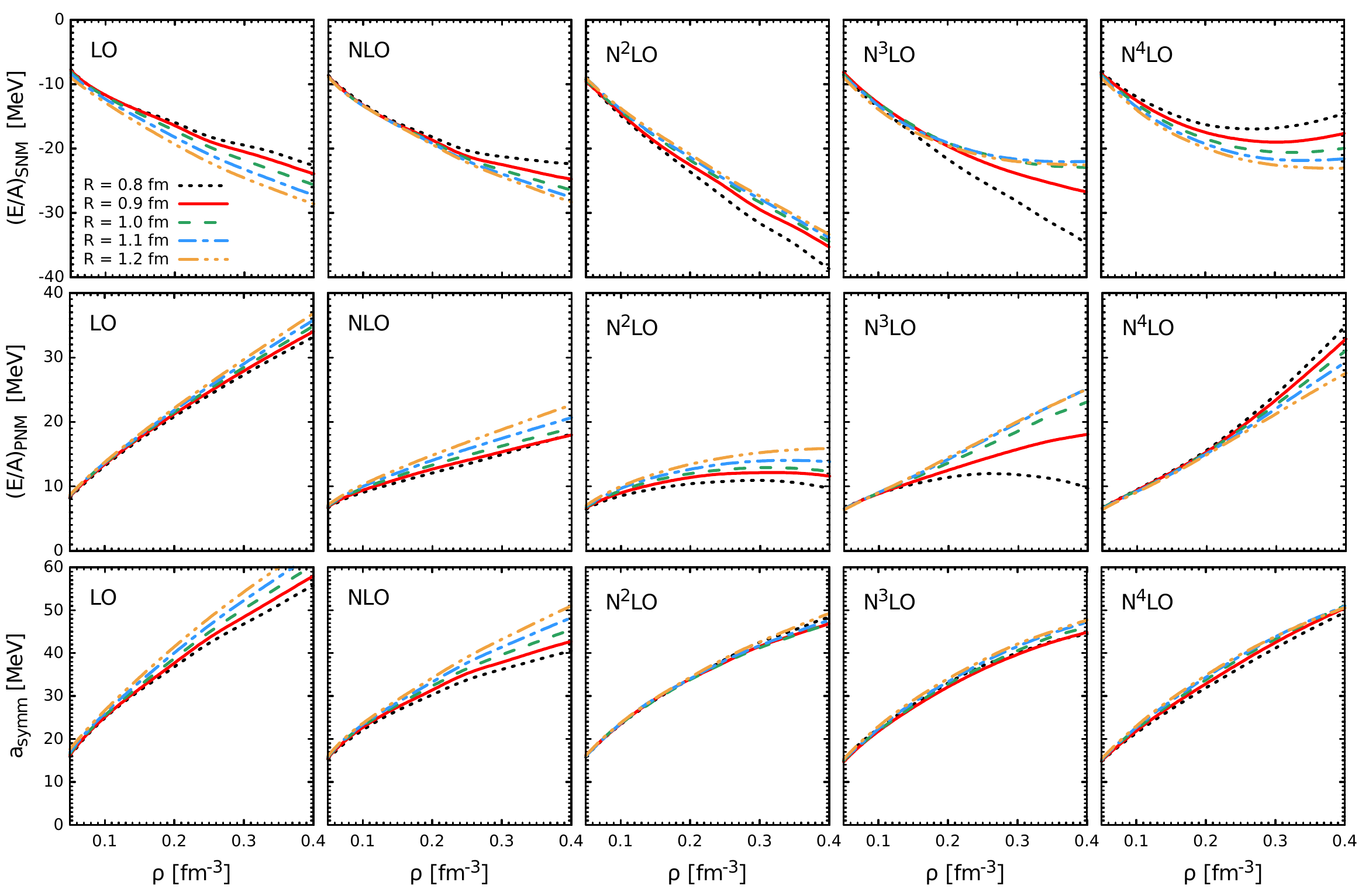}
	\caption{(Color online) Density dependence of the energy
          per particle of SNM $(E/A)_{\rm SNM}$ (upper raw), PNM
          $(E/A)_{\rm PNM}$ (middle raw) and of the
          symmetry energy $a_\text{symm}$ (lower raw) based on chiral $NN$
          potentials of \cite{epelbaum15a,epelbaum15b} for all
          available cutoff values in the range of $R=0.8\ldots 1.2~$fm.}
\label{snm}
\end{figure*}

In Fig.~\ref{snm}, we show our results for the density dependence of
the energy per nucleon of symmetric nuclear matter and pure neutron
matter for all available chiral orders and cutoff values, where the
$G-$matrices are solved up to the partial waves $J=6$.
We remind the reader that the long-range contributions are regularized in the 
newest chiral $NN$ potentials by multiplying the corresponding
coordinate-space expressions with the function 
\begin{equation}
f (r) = \bigg[ 1 - \exp \bigg( -\frac{r^2}{R^2} \bigg) \bigg]^n\,,
\quad n=6\,, \quad R = 0.8 \ldots
1.2~\mbox{fm}.   
\end{equation}
For contact interactions, a non-local Gaussian regulator in momentum space is
employed with the cutoff $\Lambda$ being related to $R$ via $\Lambda =
2/R$. We emphasize that the calculations reported in this paper do not
include the contributions of three- and four-nucleon forces and are
thus incomplete starting from N$^2$LO. 

For SNM, the LO (i.e.~$Q^0$), NLO (i.e.~$Q^2$) and N$^4$LO $NN$ potentials yield lager binding energies
for softer interactions (i.e.~for larger cutoffs $R$), while the
situation is opposite at N$^2$LO and N$^3$LO. For PNM, the harder
(softer) interactions yield more (less) attraction at
LO$\ldots$N$^3$LO (N$^4$LO). This complicated pattern suggests that the EOS is
rather sensitive to the details of the nuclear force and especially to
the interplay between its intermediate and short-range components
which is expected to be strongly regulator dependent. Our results at
NLO agree well with the ones reported in~\cite{sammarruca15} both for
SNM and PNM\footnote{We cannot compare our N$^2$LO and N$^3$LO
  predictions with those of~\cite{sammarruca15} since no results based
  on $NN$ interactions only are provided in that work.} and with the
Quantum Monte Carlos calculation of \cite{Gezerlis:2013ipa} 
for PNM. Interestingly, the cutoff dependence of the energy per
particle of PNM at NLO is qualitatively different from the one found
in \cite{sammarruca15} which demonstrates that the form of the
regulator does significantly affect the properties of the resulting potentials. 

Generally, our results for both SNM and PNM show an increasing
attraction in the $NN$ force when going from LO to N$^2$LO, that 
can probably be traced back to the two-pion exchange potential (TPEP), which
has a very strong attractive central isoscalar piece. At N$^3$LO, the
chiral TPEP receives further attractive contributions but also
develops a repulsive short-range core. The additional repulsion at
N$^4$LO comes
from   the contributions to the TPEP at
this order. The EOSs based on the N$^3$LO and N$^4$LO potentials alone
show saturation points below $\rho=0.4~$fm$^{-3}$ except for N$^3$LO at $R=0.8~$fm and $R=0.9~$fm.

\begin{table*}[!ttb]
\caption{Saturation properties of SNM based on the AV18 potential
  and the N$^4$LO chiral $NN$ potentials for all available cutoff
  values. }\label{sat}
\vskip 0.1 true cm
			\begin{tabularx}{\textwidth}{X R R R R R R}
				\hline\hline
&&&&&& \\[-10pt]
				{   }&    { AV18 }   & {N$^4$LO$_{R=0.8 \, \rm fm}$}   &{N$^4$LO$_{R=0.9 \, \rm fm}$}   & {N$^4$LO$_{R=1.0 \, \rm fm}$}   & {N$^4$LO$_{R=1.1 \, \rm fm}$}   & {N$^4$LO$_{R=1.2 \, \rm fm}$}  \\[1pt]
				\hline
&&&&&& \\[-10pt]
				{$\rho_{\text{sat}}$ (fm$^{-3})$} &  {$0.26$}    & {$0.28$}   &  {$0.29$}      & {$0.31$}           & {$0.35$}            & {$ 0.40$}       \\[1pt]
				{$E/A$ (MeV)}                     &  {$-17.78$}  & {$-17.14$} &  {$-19.15$}    & {$-20.67$}         & {$-21.92$}          & {$-23.28$}       \\[1pt]
				{$M^*/M$}                         &  {$0.71$}    & {$0.74$}   &  {$0.73$}      & {$0.72$}           & {$0.72$}            & {$0.71$}        \\[1pt]
				\hline\hline
			\end{tabularx}
\end{table*}
\begin{table*}[!htb]
\caption{Contributions of the various partial waves  (in units of MeV) to the binding energies of SNM at the corresponding saturation densities for the AV18 and chiral N$^4$LO $NN$ potentials for all available cutoff
  values.}\label{pc}
\vskip 0.1 true cm
\begin{tabularx}{\textwidth}{X R R R R R R}
					\hline\hline
&&&&&& \\[-10pt]
					{   }&    { AV18 }   & {N$^4$LO$_{R=0.8 \, \rm fm}$}  & {N$^4$LO$_{R=0.9 \, \rm fm}$}  & {N$^4$LO$_{R=1.0 \, \rm fm}$} & {N$^4$LO$_{R=1.1 \, \rm fm}$}  & {N$^4$LO$_{R=1.2 \, \rm fm}$} \\[1pt]
					\hline
&&&&&& \\[-10pt]
					{$^1S_0$} &  {$-20.71$}    & {$-18.97$}          &  {$-20.22$}         & {$-21.41$}         & {$-23.03$}          & {$-24.68$}       \\[1pt]
					{$^3P_0$} &  {$-4.74$}     & {$-4.92$}           &  {$-5.06$}          & {$-5.31$}          & {$-5.75$}           & {$-6.21$}       \\[1pt]
					{$^3S_1$-$^3D_1$} &  {$-21.91$}    & {$-23.83$}  &  {$-24.87$}         & {$-25.80$}         & {$-26.60$}          & {$-27.27$}        \\[1pt]
					{$^3P_1$} &  {$16.68$}     & {$18.29$}           &  {$19.07$}          & {$20.62$}          & {$23.78$}           & {$27.64$}       \\[1pt]
					{$^1P_1$} &  {$6.22$}      & {$7.02$}            &  {$7.23$}           & {$7.75$}           & {$8.83$}            & {$10.16$}       \\[1pt]
					{$^3P_2$-$^3F_2$} &  {$-13.96$}    & {$-15.92$}  &  {$-16.68$}         & {$-18.17$}         & {$-21.23$}          & {$-25.25$}        \\[1pt]
					{$^1D_2$} &  {$-4.94$}     & {$-5.44$}           &  {$-5.71$}          & {$-6.29$}          & {$-7.49$}           & {$-9.05$}       \\[1pt]
					{$^3D_2$} &  {$-6.89$}     & {$-7.62$}           &  {$-7.97$}          & {$-8.66$}          & {$-10.05$}          & {$-11.78$}       \\[1pt]
					{$^3D_3$$-^3G_3$} &  {$0.25$}      & {$0.49$}    &  {$0.46$}           & {$0.43$}           & {$0.39$}            & {$0.28$}        \\[1pt]
					{$^1F_3$} &  {$1.44$}      & {$1.55$}            &  {$1.62$}           & {$1.76$}           & {$2.03$}            & {$2.37$}       \\[1pt]
					{$^3F_3$} &  {$2.65$}      & {$2.93$}            &  {$3.06$}           & {$3.35$}           & {$3.90$}            & {$4.58$}       \\[1pt]
					{$^3F_4$-$^3H_4$} &  {$-1.02$}     & {$-1.26$}   &  {$-1.32$}          & {$-1.46$}          & {$-1.76$}           & {$-2.12$}         \\[1pt]  
					{$^1G_4$} &  {$-0.88$}      & {$-1.00$}          &  {$-1.06$}          & {$-1.17$}          & {$-1.41$}           & {$-1.72$}       \\[1pt]
					{$^3G_4$} &  {$-1.47$}      & {$-1.65$}          &  {$-1.74$}          & {$-1.93$}          & {$-2.34$}           & {$-2.87$}       \\[1pt]
					\hline\hline
\end{tabularx}
\end{table*}

\begin{table*}[!htb]
	\caption{Contributions of the various partial waves  (in units of
		MeV) to the binding energies of SNM at the empirical saturation density, $\rho=0.16$ fm$^{-3}$, 
		for the AV18 and chiral N$^4$LO $NN$ potentials for all available cutoff
		values.}\label{pc1}
	\vskip 0.1 true cm
	\begin{tabularx}{\textwidth}{X R R R R R R}
		\hline\hline
		&&&&&& \\[-10pt]
		{   }&    { AV18 }   & {N$^4$LO$_{R=0.8 \, \rm fm}$}  & {N$^4$LO$_{R=0.9 \, \rm fm}$}  & {N$^4$LO$_{R=1.0 \, \rm fm}$} & {N$^4$LO$_{R=1.1 \, \rm fm}$}  & {N$^4$LO$_{R=1.2 \, \rm fm}$} \\[1pt]
		\hline
		&&&&&& \\[-10pt]
		{$^1S_0$} &  {$-15.01$}    & {$-14.32$}          &  {$-14.83$}         & {$-15.19$}         & {$-15.47$}          & {$-15.81$}       \\[1pt]
		{$^3P_0$} &  {$-3.07$}     & {$-3.17$}           &  {$-3.17$}          & {$-3.18$}          & {$-3.18$}           & {$-3.18$}       \\[1pt]
		{$^3S_1$-$^3D_1$} &  {$-18.74$}    & {$-19.72$}  &  {$-20.18$}         & {$-20.68$}         & {$-20.78$}          & {$-20.93$}        \\[1pt]
		{$^3P_1$} &  {$8.47$}      & {$9.16$}            &  {$9.17$}           & {$9.14$}           & {$9.15$}            & {$9.14$}       \\[1pt]
		{$^1P_1$} &  {$3.36$}      & {$3.61$}            &  {$3.59$}           & {$3.57$}           & {$3.56$}            & {$3.55$}       \\[1pt]
		{$^3P_2$-$^3F_2$} &  {$-6.89$}    & {$-7.71$}    &  {$-7.71$}          & {$-7.73$}          & {$-7.74$}           & {$-7.79$}        \\[1pt]
		{$^1D_2$} &  {$-2.26$}     & {$-2.45$}           &  {$-2.45$}          & {$-2.47$}          & {$-2.50$}           & {$-2.55$}       \\[1pt]
		{$^3D_2$} &  {$-3.34$}     & {$-3.65$}           &  {$-3.65$}          & {$-3.66$}          & {$-3.67$}           & {$-3.68$}       \\[1pt]
		{$^3D_3$$-^3G_3$} &  {$0.08$}      & {$0.20$}    &  {$0.19$}           & {$0.16$}           & {$0.13$}            & {$0.09$}        \\[1pt]
		{$^1F_3$} &  {$0.66$}      & {$0.72$}            &  {$0.72$}           & {$0.72$}           & {$0.72$}            & {$0.72$}       \\[1pt]
		{$^3F_3$} &  {$1.19$}      & {$1.31$}            &  {$1.30$}           & {$1.30$}           & {$1.30$}            & {$1.29$}       \\[1pt]
		{$^3F_4$-$^3H_4$} &  {$-0.34$}     & {$-0.41$}   &  {$-0.41$}          & {$-0.40$}          & {$-0.39$}           & {$-0.38$}         \\[1pt]  
		{$^1G_4$} &  {$-0.35$}      & {$-0.39$}          &  {$-0.39$}          & {$-0.39$}          & {$-0.39$}           & {$-0.39$}       \\[1pt]
		{$^3G_4$} &  {$-0.57$}      & {$-0.64$}          &  {$-0.64$}          & {$-0.63$}          & {$-0.63$}           & {$-0.63$}       \\[1pt]
		\hline\hline
	\end{tabularx}
\end{table*}

It is instructive to compare the results based on the most accurate
chiral potentials at N$^4$LO with the ones from high-precision
phenomenological interactions such as the AV18 potential \cite{wiringa95}.  
In Table.~\ref{sat}, we list the saturation properties, saturation densities and saturation binding energies per particle, and the
effective mass of the nucleon \cite{li16}:
\beq
\frac{M^*}{M}=1-\frac{dU(k;e(k))}{de(k)},
\eeq
at the saturation point for the AV18 and N$^4$LO potentials, while the contributions of the
various partial waves up to $J=4$ to the potential energy per nucleon at the saturation
density are given in  Table~\ref{pc}. Notice that the listed
saturation properties are still far from the empirical data
($\rho_{\text{sat}}\sim0.16$~fm$^{-3}$ and $E/A\sim 16$~MeV) due to
the missing 3NF contributions~\cite{li06,baldo07}. Naturally, we
observe that the results based on the hardest version of the N$^4$LO potential with
$R=0.8~$fm are rather similar to those based on AV18. Interestingly,
we find that the partial wave contributions to the energy 
increase when the N$^4$LO potentials are softened by increasing the
coordinate-space cutoff $R$ (except for the $^3D_3$-$^3G_3$ channel). In Table \ref{pc1}, the partial wave contributions to potential energy at the empirical saturation density, $\rho=0.16$ fm$^{-3}$ for different $NN$ potentials are listed from $^1S_0$ to $^3G_4$ states. It is found that all contributions are nearly cutoff-independent expect the ones from $^1S_0$, $^3S_1$-$^3D_1$, and $^3D_3$-$^3G_3$ states, which are decreasing with the cutoffs $R$. Actually, the size of these contributions is strongly dependent on the central and tensor components in the $NN$ potential. The larger cutoff $R$ corresponds to stronger short-range correlations and removes more repulsive contribution on the $NN$ potential at short distance. It will generate more attractive binding energy. Our results for the saturation density and binding energy confirm the
linear correlation between these two quantities known as the Coester
line \cite{Coester:1970ai}, see also \cite{li06}.  
Calculations within the BHF theory using phenomenological potentials
have revealed that the position on the Coester line is 
correlated with the deuteron $D$-state probability $P_D$ with 
smaller values of $P_D$ typically resulting in  smaller 
saturation energy and density~\cite{machleidt89,li06}. 
We do not observe this correlation for the chiral N$^4$LO potentials 
with $P_D=4.28\%$ ($P_D=5.12\%$) for $R=0.8~$fm ($R=1.2~$fm).
This is similar to the lack of correlation between
$P_D$ and the triton binding energy for the novel chiral potentials \cite{binder16}. 
We remind the reader that the $D$-state probability is not an observable.

We have also extracted the symmetry energy of
nuclear matter $a_\text{symm} (\rho )$,
the quantity which 
 describes the response of the nuclear force on excess neutrons or
 protons and plays an important role in understanding the properties
 of nuclei and astrophysical objects. The symmetry energy $a_\text{symm} (\rho ) $ is defined
 in terms of the expansion of the asymmetric nuclear matter in 
 powers of the asymmetry parameter $\delta \equiv (\rho_n - \rho_p )/
 \rho $, with $\rho_n$ and  $\rho_p$ referring to the neutron and
 proton number densities, via
\begin{equation}
\frac{E}{A} (\rho, \delta ) =  \frac{E}{A} (\rho, 0 ) + a_\text{symm}
(\rho ) \,  \delta^2 + \ldots \,.
\end{equation}
The terms beyond the quadratic one are known to be very small
\cite{Drischler:2015eba},  so that
the symmetry energy can be well approximated by 
\begin{equation}
a_\text{symm} (\rho ) = \left( \frac{E}{A} \right)_{\rm PNM} - \left( \frac{E}{A} \right)_{\rm SNM} \,,
\end{equation}
where $E/A$ is viewed as a function of $\rho$ and $\delta$. 
While the calculated symmetry energies show significant cutoff
dependence at LO and NLO, which is comparable to that of $(E/A)_{\rm SNM}$ and
$(E/A)_{\rm PNM}$, the results at higher orders are almost insensitive
to the values of $R$ and show a little variation with the order of the
chiral expansion. The resulting value of  $a_\text{symm}
=27.9-30.5~$MeV at the empirical saturation density, calculated using the N$^4$LO potentials, is consistent with the
empirical constraints and the results from the phenomenological high-precision $NN$ potentials \cite{li06} with $a_\text{symm}
	=28.5-32.6~$MeV at $\rho=0.17$ fm$^{-3}$ and the ones from the functional renormalization group method with $a_\text{symm}
	=29.0-33.0~$MeV at $\rho=0.16$ fm$^{-3}$ \cite{drews15}. Furthermore, Vida\~na {\it{et al.}} also studied the properties of the symmetry energy with AV 18 potential plus a
phenomenological three-body force as Urbana type \cite{vidana09}. However, it is found that the isovector properties of nuclear matter are not affected by the three-body force too much.

\section{Uncertainty quantification}
\label{sec:Errors}

\begin{figure}[tb]
	\centering
	\includegraphics[width=0.48\textwidth]{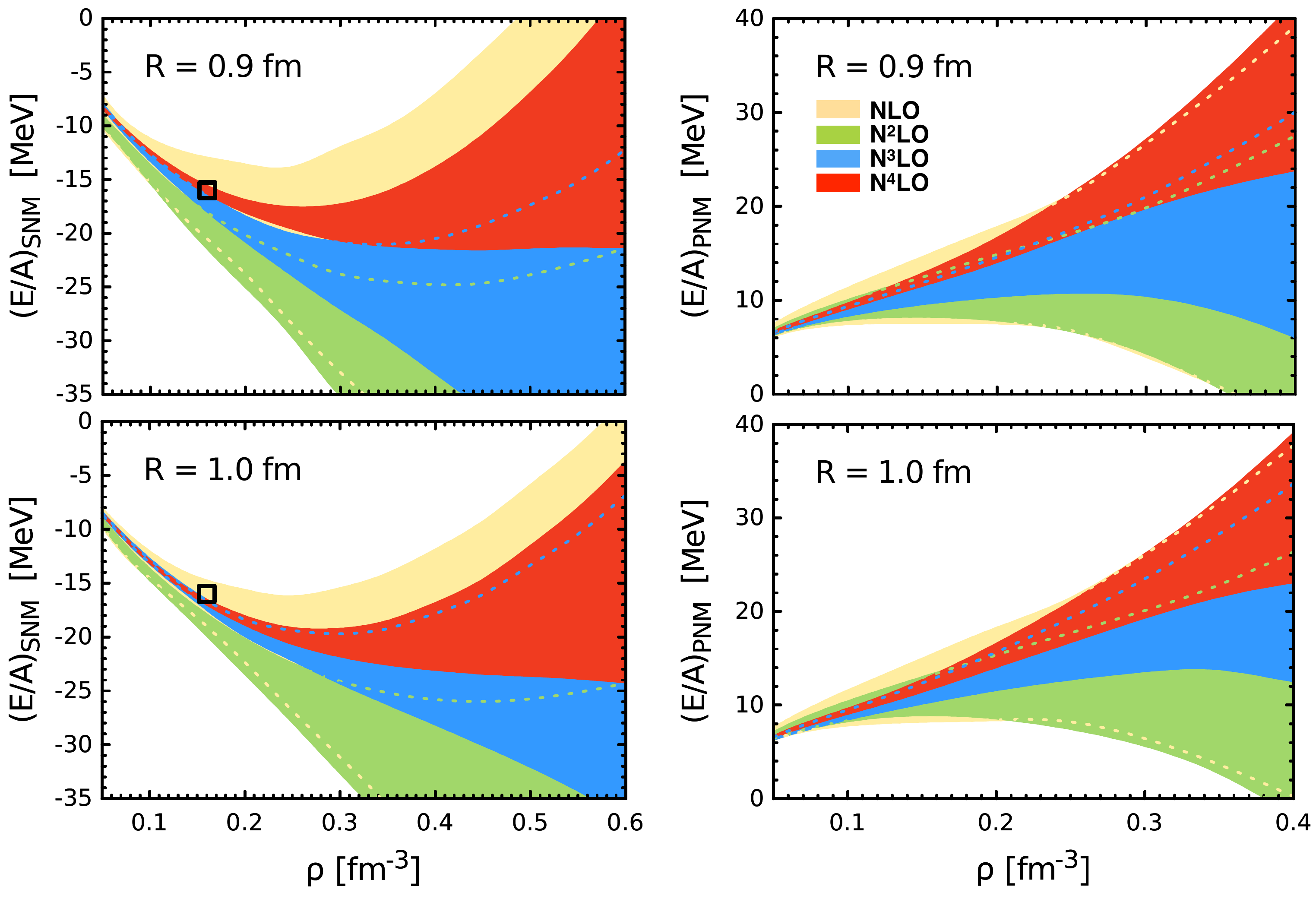}
\vskip -0.2 true cm
	\caption{(Color online) Predictions for the EOS of SNM (left
          panel) and PNM (right panel)
          based on the chiral $NN$ potentials of
          Refs.~\cite{epelbaum15a,epelbaum15b} for $R=0.9~$fm (upper
          raw) and $R=1.0~$fm (lower raw) along with the
          estimated theoretical uncertainties. Open rectangles
          visualize the empirical saturation point of symmetric
          nuclear matter.  }\label{NMErrors}
\end{figure}

We now turn to the important question of uncertainty quantification from the truncation of the chiral expansion. Actually, Baldo {\it{et al.}} attempted to quantify the theoretical uncertainties of the EOSs with the family of Argonne $NN$ potential through comparing the BHF theory to other many-body approaches \cite{baldo12}. These uncertainties are strongly dependent on the methodologies of nuclear many-body approximation to treat the spin structures of potentials.  Here we follow the approach formulated in Ref.~\cite{epelbaum15a}, which makes use of the explicitly
known contributions to an observable of interest at various chiral
orders to estimate the size of truncated terms \emph{without relying
  on cutoff variation}. The algorithm proposed
in \cite{epelbaum15a} has been adjusted in Ref.~\cite{binder16} to
enable applications to incomplete few- and many-nucleon calculations
based on two-nucleon
forces only.  Here and in what follows, we use the method as
formulated in that paper, which was also employed in \cite{Skibinski:2016dve}. Specifically, for an observable $X (p)$ with
$p$ referring to the corresponding center-of-mass momentum scale, the 
theoretical uncertainty $\delta X^{(i)}$ of the $i$-th chiral order
prediction $X^{(i)}$  is estimated via 
\begin{eqnarray}
\delta X^{(0)} &=& \max (Q^2 | X^{(0)}|, \, |X^{(\geq 0)} -X^{(\geq
  0)}| ), \nonumber \\
\delta X^{(2)} &=& \max (Q^3 | X^{(0)}|, \,Q | \Delta X^{(2)} |, \, Q
                   \delta X^{(0)}, \,  |X^{(\geq 2)} -X^{(\geq
  2)}|), \nonumber \\
\delta X^{(i)} &=& \max (Q^{i+1} | X^{(0)}|, \,Q^{i-1} | \Delta
                   X^{(2)} |, \, Q^{i-2} | \Delta X^{(3)} |,  Q \delta X^{(i-1)}) \quad \mbox{for} \;  i \geq 3\,,
\end{eqnarray}
where $Q = \max ( p/\Lambda_{\rm b}, \; M_\pi/\Lambda_{\rm b} )$ is
the estimated expansion parameter while $\Delta X^{(2)} \equiv X^{(2)}
- X^{(0)}$ and $\Delta X^{(i)} \equiv X^{(i)}
- X^{(i-1)}$,  $i> 2$,  denote the chiral-order $Q^2$ and $Q^i$ contributions
to $X(p)$. The breakdown scale of the nuclear chiral EFT was 
estimated to be $\Lambda_{\rm b} \simeq 600~$MeV \cite{epelbaum15a}.\footnote{To account for increasing finite-cutoff artefacts using softer
versions of the chiral forces, the lower values of $\Lambda_{\rm b} = 500~$MeV
and $400~$MeV were employed in calculations based on $R=1.1~$fm and 
$R=1.2~$fm, respectively.} The Bayesian analysis of the chiral EFT
predictions for the $NN$ total cross section of
Ref.~\cite{Furnstahl:2015rha} has revealed, 
that the actual breakdown scale may even be a little higher
than $\Lambda_{\rm b} \simeq 600~$MeV for $R=0.9~$fm.

\begin{figure}[tb]
	\centering
	\includegraphics[width=0.48\textwidth]{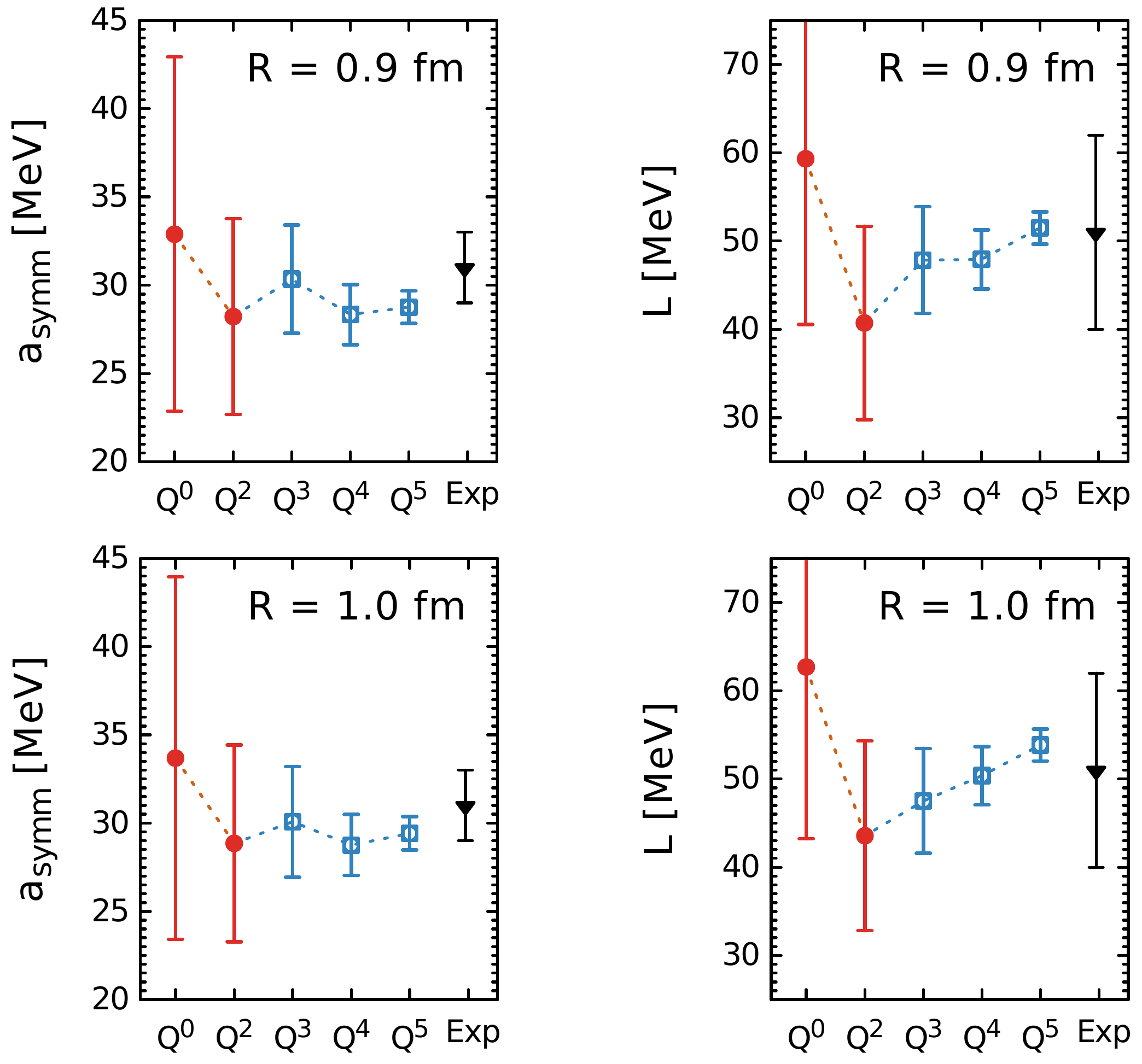}
	\caption{(Color online) Chiral expansion of the symmetry
          energy $a_\text{symm}$ (left panel) and the slope parameter
          $L$ (right panel) at the empirical saturation density of $\rho=0.16$ fm$^{-3}$ for the cutoff values of $R=0.9~$fm (upper
          raw) and $R=1.0~$fm (lower raw) along with the estimated theoretical
          uncertainty. Solid circles (open rectangles) show the complete results at a given chiral order  
          (incomplete results based on $NN$ interactions only).  Solid triangles show the current experimental constraints on
          $a_\text{symm}$  and $L$ as described in the text.}\label{SLErrors}
\end{figure}

In Fig.~\ref{NMErrors}, we show the results for the EOS for SNM and PNM including the
estimated theoretical uncertainties at various orders of the chiral
expansion for the most accurate versions of the $NN$ potentials
with $R=0.9~$fm and $R=1.0$~fm \cite{epelbaum15a,epelbaum15b}. The expansion parameter $Q$ at a given
density is estimated by identifying the momentum scale $p$ with the Fermi
momentum $k_{\rm F}$, which is related to the density $\rho$ via $\rho
= 2 k_{\rm F}^3 /(3 \pi^2)$ ($\rho = k_{\rm F}^3 /(3 \pi^2)$) for SNM
(PNM), and assuming $\Lambda_{\rm b} = 600~$MeV. At the saturation density, the achievable accuracy of the chiral EFT
predictions for the energy per particle may be expected to be about 
$\pm 1.5~$MeV ($\pm 0.3~$MeV) for SNM and $\pm 2~$MeV ($\pm 0.7~$MeV)
for PNM at N$^2$LO (N$^4$LO).  Notice that the expected accuracy at
N$^4$LO is significantly smaller than the current model dependence for
these quantities. We further emphasize that the presented estimations should be
taken with some care due to the non-availability of complete calculations beyond
NLO. More reliable estimations of the theoretical uncertainty using
the approach of \cite{epelbaum15a} will be possible once the
corresponding three- and four-nucleon forces are included. 

Our results confirm the conclusions of \cite{sammarruca15} that cutoff
variation does not provide an adequate way for estimating the
uncertainties in the calculations of the nuclear EOS. As discussed in
\cite{epelbaum15a}, the residual cutoff-dependence of observables may
generally be expected to underestimate the theoretical uncertainty at NLO
and N$^3$LO, which is consistent with our results. Further,
the spread of results for different values of $R$ at N$^4$LO appears
to be roughly of a similar size as the estimated uncertainty at this order. 
We, however, refrain from drawing more definite conclusions on the cutoff
dependence based on the incomplete calculations. 

Finally, we have also quantified the achievable accuracy of the theoretical
determination of the symmetry energy $a_\text{symm}$ and the slope
parameter $L$, defined as $L = 3 \rho \, \partial (E/A)_{\rm
  SNM}/\partial \rho$, at the empirical saturation
density. These important quantities have been constrained by the
available experimental information on e.g.~neutron skin thickness, heavy
ion collisions and dipole polarizabilities leading to the ranges 
of $29\mbox{ MeV} \lesssim a_\text{symm} \lesssim 33\mbox{ MeV}$ and 
$40\mbox{ MeV} \lesssim L \lesssim 62\mbox{ MeV}$ \cite{Tsang:2012se,Lattimer:2012xj,Lattimer:2014sga}. In
Fig.~\ref{SLErrors}, we show our results for these quantities using the $NN$
potentials from LO to N$^4$LO along with the estimated theoretical
uncertainties. Especially for the slope parameter, a complete
calculation at N$^4$LO would yield a theoretical prediction much
more accurate than the current experimental data.

\section{Summary and conclusions}
\label{sec:Summary}

In summary, we calculated the equations of state (EOSs) of SNM and PNM 
with the state-of-the-art chiral $NN$ potentials from LO to N$^4$LO 
in the framework of Brueckner-Hartree-Fock
theory. At N$^4$LO, the  EOS of SNM has saturation points for all
employed cutoff values with the corresponding saturation densities and
binding energies per particle being within the range of $0.28\ldots 0.40$ fm$^{-3}$
and $-17.14 \ldots -23.28~$MeV, respectively. These values are compatible
with the ones based on the phenomenological high-precision potentials
like  e.g.~the AV18 potential. The symmetry
energy and the slope parameter at the saturation density are found to
be in the range of $a_\text{symm} = 27.9 \ldots 30.5~$MeV and 
 $L = 49.4 \ldots 55.0~$MeV, respectively, using the N$^4$LO potentials with the
 cutoff in the range of $R=0.8\ldots 1.2~$fm.  

We have also estimated the achievable theoretical accuracy at various
orders in the chiral expansion using the novel approach formulated in
Refs.~\cite{epelbaum15a,binder16} and discussed the convergence of the
chiral expansion. Similar to \cite{sammarruca15}, we find that
the residual cutoff dependence of the energy per particle does not
allow for a reliable estimation of the theoretical uncertainty, see also the
discussion in Ref.~\cite{epelbaum15a}.  We find that chiral EFT may be
expected to provide an accurate description of  SNM and PNM at the 
saturation density, with the expected accuracy of a few
percent at N$^4$LO. At this order, a semi-quantitative description of
the EOS should be possible up to about twice the saturation
density of nuclear matter. Clearly, this will require a consistent
inclusion of the corresponding many-body forces. Work along these
lines is in progress to compare with the existing calculations with two-body and three-body chiral force \cite{carbone14,sammarruca15}.

\section*{Acknowledgments}
We would like to thank Arnau Rios Huguet for sharing his insights into the topics discussed here.
UGM thanks the ITP (CAS, Beijing) for hospitality, where part of this work was done. 
This work was supported in part by the National Natural Science Foundation of China (Grant Nos. 11335002, 11405090, 11405116 and 11621131001), DFG (SFB/TR 110, ``Symmetries and the Emergence of Structure in QCD'') and BMBF (contract No. 05P2015 -NUSTAR R\&D). The work of UGM was supported in part by The Chinese Academy of Sciences 
(CAS) President's International Fellowship Initiative (PIFI) grant no. 2015VMA076.


\end{document}